\documentclass{aa}
\usepackage[varg]{txfonts}

\bibpunct{(}{)}{;}{a}{}{,} 

\usepackage{graphicx}

\begin{document}

   \title{Nanoflare distributions over solar cycle 24 based on SDO/AIA differential emission measure observations}


   \author{Stefan Purkhart
          \inst{1}
          \and
          Astrid M. Veronig\inst{1}\fnmsep\inst{2}
          }

   \institute{Institute of Physics, University of Graz, Universitätsplatz 5, 8010 Graz, Austria\\
              \email{stefan.purkhart@uni-graz.at}
         \and
             Kanzelhöhe Observatory for Solar and Environmental Research, University of Graz, Kanzelhöhe 19, 9521 Treffen, Austria\\
             \email{astrid.veronig@uni-graz.at}
             }

   \date{Received ; accepted }

 
  \abstract
   {}
   {Nanoflares in quiet-Sun regions during solar cycle 24 are studied with the best available plasma diagnostics to derive their energy distribution and contribution to coronal heating during different levels of solar activity.}
   {Extreme ultraviolet (EUV) filters of the Atmospheric Imaging Assembly (AIA) onboard the Solar Dynamics Observatory (SDO) are used. We analyze 30 AIA/SDO image series between 2011 and 2018, each covering a $400\arcsec\times 400\arcsec$ quiet-Sun field-of-view over two hours with a 12-second cadence. Differential emission measure (DEM) analysis is used to derive the emission measure (EM) and temperature evolution for each pixel. We detect nanoflares as EM-enhancements using a threshold-based algorithm and derive their thermal energy from the DEM observations.}
   {Nanoflare energy distributions follow power-laws that show slight variations in steepness ($\alpha =$ 2.02 to 2.47) but no correlation to the solar activity level. The combined nanoflare distribution of all data sets covers five orders of magnitude in event energies ($10^{24} \mathrm{~to~} 10^{29} \mathrm{~erg}$) with a power-law index $\alpha=2.28 \pm0.03$. The derived mean energy flux of $(3.7\pm 1.6)\times 10^4\mathrm{~erg~cm^{-2}~s^{-1}}$ is one order of magnitude smaller than the coronal heating requirement. We find no correlation between the derived energy flux and solar activity. Analysis of the spatial distribution reveals clusters of high energy flux (up to $3\times 10^5 \mathrm{~erg~cm^{-2}~s^{-1}}$) surrounded by extended regions with lower activity. Comparisons with magnetograms from the Helioseismic and Magnetic Imager (HMI) demonstrate that high-activity clusters are located preferentially in the magnetic network and above regions of enhanced magnetic flux density.}
   {The steep power-law slope ($\alpha>2$) suggests that the total energy in the flare energy distribution is dominated by the smallest events, i.e., nanoflares. We demonstrate that in the quiet Sun, the nanoflare distributions and their contribution to coronal heating does not vary over the solar cycle.}

   
   \titlerunning{Nanoflare distributions over solar cycle 24}
	\authorrunning{Stefan Purkhart \& Astrid M. Veronig}

   \maketitle
%

\section{Introduction}\label{sec:introduction}

The temperature of the solar corona of several million Kelvin is one of the most challenging mysteries of solar physics. Ever since the discovery of the forbidden lines of highly ionized iron atoms in the coronal spectrum by \citet{Grotrian1939} and \citet{Edlen1943}, many studies have been carried out to solve the resulting \emph{coronal heating problem}.
The multi-million Kelvin coronal temperature results in heat losses of at least $3\times 10^{5}\mathrm{~erg~cm^{-2}~s^{-1}}$ \citep{Withbroe1977}, which must be continuously balanced by energy input to prevent cooling and collapsing of the corona. However, the heat flux from the lower atmospheric layers cannot provide the required energy because the temperatures there are much lower. Consequently, some processes have to convert a non-thermal energy source into thermal energy in order to maintain the coronal plasma at stable million-Kelvin temperatures. Several models have been developed and extensively tested, and they provide us with at least part of the explanation. They can basically be assigned into two groups, which explain coronal heating either by various types of magnetized waves that are initiated by photospheric plasma motions, propagating upward into the corona and dissipating their energy there or by small-scale magnetic reconnection events dubbed \emph{nanoflares} \citep[e.g., reviews by][]{Klimchuk2006,Parnell2012}.

Nanoflares were foreseen by \citet{Parker1988_Nanoflares} and could provide a solution to the coronal heating problem if they occur at a sufficiently high rate. They are the result of photospheric granular and supergranular motions that move the base of field lines in a random-walk-like manner and result in field lines that are twisted and wrapped around each other \citep{Parker1983}. Current sheets form at the interface of misaligned flux tubes and build up free energy, which is then spontaneously released by magnetic reconnection and partially converted into thermal energy. Self-organized criticality models can explain this process of accumulation until a critical state followed by a subsequent release \citep[][]{Lu1991}. Here, regions of all sizes can be on the verge of stability and produce reconnection avalanches that we observe as flares over many orders of magnitude in event energy.

It has been shown that the occurrence frequency $(dN/dE)$ of solar flares generally follows a power-law distribution \citep[e.g.][]{Dennis1985,Crosby1993,Veronig2002,Hannah2008}, in line with the scale-invariance expected from self-organized criticality models. The power-law can be expressed as a function of flare energy ($E$), a power-law index ($\alpha$), and a normalization factor ($A$):
\begin{equation}
  \frac{dN}{dE} = AE^{-\alpha}
  \label{equ:freqdist}
\end{equation}
Integration over the whole energy range (from $E_{min}$ to $E_{max}$) of flares gives the total energy released:

\begin{equation}
  W(E_{min} \leq E \leq E_{max}) = \int_{E_{min}}^{E_{max}} \frac{dN}{dE} E dE = \frac{A}{2-\alpha} [E_{max}^{-\alpha +2}-E_{min}^{-\alpha +2}]
\end{equation}\\
\citet{Hudson1991} first pointed out that the relative contribution of small and large flares to the total energy in such a power-law distribution depends critically on the power-law index $\alpha$. In the case $\alpha>2$, the lower energy range of the frequency distribution dominates the total energy input, making it possible for nanoflares to provide sufficient energy for coronal heating.
Observations of regular flares have shown that they cannot provide enough energy and that their power-law index is less than 2 \citep{Crosby1993}.
In order to explain coronal heating through magnetic reconnection, the nanoflare distribution may have to become steeper than the common power-law observed for large flares \citep{Lu1991}.
However, it has to be noted that the relative contribution alone cannot be used to determine whether the sum of all events provides enough energy.

Nanoflares are the smallest flare-like energy releases on the Sun. While regular flares and microflares occur only in Active Regions \citep[e.g.][]{Stoiser2007,Hannah2011,Fletcher2011}, nanoflares occur also in quiet-Sun regions \citep[e.g.][]{Krucker1998,Parnell2000} and are thus a possible means to serve for the heating of the quiet (non-AR) corona.
Several studies have attempted to determine the frequency distribution of nanoflares and their contribution to coronal heating.
Early observational nanoflares studies determined the nanoflare energies from the emission measure and temperature calculated using the spectral line ratio method applied to two EUV filters. Data was used either from the EUV Imaging Telescope (EIT) \citep{Delaboudiniere1995} onboard the Solar and Heliospheric Observatory (SOHO) or the Transition Region and Coronal Explorer (TRACE) \citep{Handy1999}. These nanoflares studies \citep{Krucker1998,Parnell2000,Aschwanden2000_1,Krucker2002,Aschwanden2002} found power-laws in the range of 1.8 to 2.7 that allowed no definitive conclusion on whether small or large flares contribute more energy to the coronal heating. The used instruments also only allow rudimentary event energy calculation because of the limited temperature coverage of the imaged wavelengths. In a more recent study, \citet{Joulin2016} use the Atmospheric Imaging Assembly (AIA) aboard the Solar Dynamics Observatory (SDO) because of its much better plasma diagnostics and arrive at a power-law index of 1.65 and 1.73 for the thermal event energy with background removal for an active and quiet-Sun region, respectively. They were, however, limited to pre-calculated DEMs from \citet{Guennou2012_2} that lack the full AIA temporal resolution.

The listed nanoflare studies used different instruments with different spatial and temporal resolution and diagnostic capabilities. They further used different event detection methods, energies were calculated as thermal energy or radiative/conductive losses, and they also varied in other assumptions. Lastly, they focused on regions of different solar activity, ranging from quiet-Sun regions to active regions, and were done at different times during the solar cycle. While all studies provide us with important information about nanoflares distributions in different solar regions, they are hard to compare, and we can not conclude how observations of similar regions might change when done at different times during the solar cycle.

By observing nanoflares in quiet-Sun regions throughout different phases of the solar cycle with the same instrument and the same detection algorithm, we can investigate about systematic changes to the slope of the power-law, the energy input into the corona, and other nanoflare parameters and check their correlation to the level of solar activity. Any changes obtained, whether they are actual variations of the observed nanoflares or variations induced by the instrument and image properties (e.g., count rate and contrast changing due to degradation), can aid the design of future studies and help compare previous studies with each other.

Since its launch in 2010, the Atmospheric Imaging Assembly (AIA) onboard the Solar Dynamics Observatory (SDO) has captured full disk, high spatial resolution solar images with a high cadence at multiple EUV wavelengths. The available data, therefore, covers a significant portion of solar cycle 24. Filtergrams are recorded at 6 coronal EUV wavelengths with different temperature responses. This allows us to derive detailed plasma diagnostics from these images using Differential Emission Measure (DEM) analysis.

This paper is structured as follows. In section \ref{sec:methods} we describe the data sets and DEM analysis, the developed event detection algorithm, and the event energy calculation. The results of this analysis are presented in section \ref{sec:results}. We show the observed frequency distributions, event numbers, energy flux, spatial distribution, and event areas for all data sets and correlate them to the solar activity cycle. Furthermore, the combined frequency distribution and energy flux from all data sets are derived. Section \ref{sec:discussion} follows with a discussion of the obtained power-laws and their implications for the contribution to coronal heating of the detected events. Finally, in section \ref{sec:conclusion} we present our conclusions on the applied methods and the results obtained.

\section{Methods}\label{sec:methods}

\subsection{Data}
\label{sec:methods_data}

We use data from the Atmospheric Imaging Assembly \citep[AIA;][]{Lemen2012_AIA} onboard the Solar Dynamics Observatory \citep[SDO;][]{Pesnell2012_SDO} that orbits the Earth in a geosynchronous orbit since 11 February 2010. The AIA instrument consists of four individual f/20 Cassegrain telescopes with 20 cm primary optics, active secondary mirrors, and a CCD sensor with $4096 \times 4096$ pixels, each corresponding to a spatial sampling of the solar disk with $0.6\arcsec$. With this configuration, AIA produces full disc images of the Sun in multiple wavelengths with a $1.5\arcsec$ spatial resolution at a standard operating cadence of 12 seconds in the EUV and 24 s in the UV \citep{Lemen2012_AIA}.

\begin{figure*}
\centering
  \includegraphics[width=16.5cm]{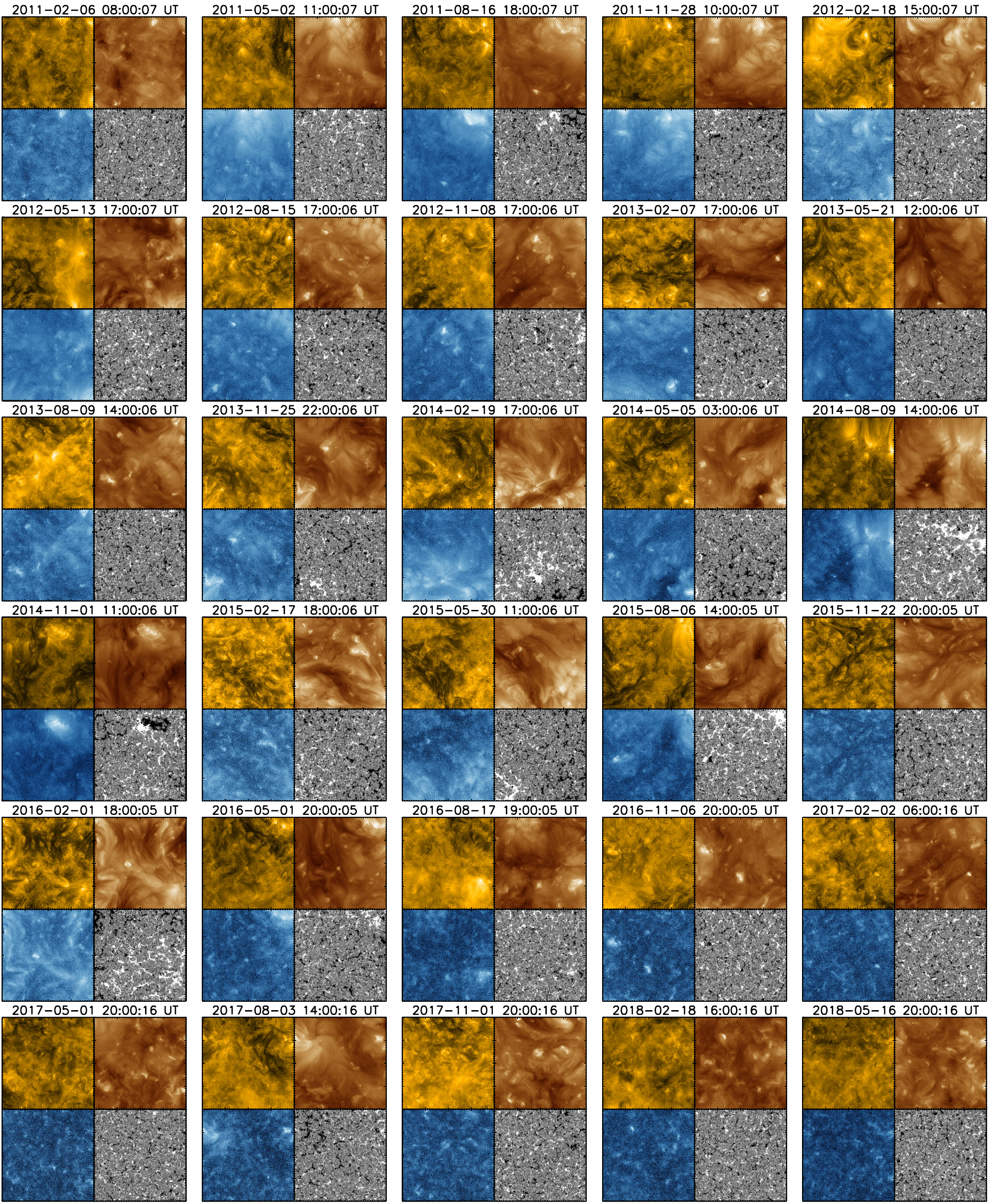}
    \caption{Groups of the first AIA 171 Å, 193 Å, and 335 Å images for each of the 30 data sets plus the HMI line-of-sight magnetogram (saturated at $\pm15$ G.) from the beginning of each observation series. All images focus on a $400\arcsec \times 400\arcsec$ FOV around the center of the solar disk. The start of the observation time is displayed above each image group, and the complete data sets consist of all images taken during the two hours following the shown images.}
    \label{fig:SolarCycleOverview}
\end{figure*}

We use images taken in the six extreme ultraviolet (EUV) channels (94, 131, 171, 193, 211, and 335 Å) that are centered on wavelengths corresponding to essential iron lines at various ionization stages (Fe\,{\sc xviii}, Fe\,{\sc viii} and Fe\,{\sc xxi}  , Fe\,{\sc ix}, Fe\,{\sc xii} and Fe\,{\sc xxiv}, Fe\,{\sc xiv}, and Fe\,{\sc xvi}). These EUV wavelength channels are not only sensitive to the line's peak formation temperatures (log T [K]: 6.8, 5.6 and 7.0, 5.8, 6.2 and 7.3, 6.3, 6.4) but show a much broader temperature response that makes the chosen EUV wavelengths sensitive to plasma temperatures in a range of at least $0.1$ to $20$ MK \citep{Lemen2012_AIA,Boerner2012}.

To obtain a uniform distribution of data sets over the solar cycle, we searched for suitable image series in February, April, August, and November of each year. All image series include an observation duration of 2 hours with the full AIA cadence of 12 seconds and focus on the center of the solar disk to minimize influences from projection effects and to have similar observation conditions for all sets. To select the data sets, beginning on the first day of the selected month, image series with the desired duration and field-of-view (FOV) were selected based on the following criteria: No active regions or coronal holes in the considered FOV and no flares during the two-hour observation period. The exclusion of active regions and coronal holes ensures that we only include regions of the quiet Sun in the analysis. Excluding data with simultaneous flares (even outside the considered FOV) prevents the use of short-exposure AIA images, which would result in low counts in the quiet-Sun regions and consequently poor DEM performance. If no such image series could be found for the first day of the month, the search was continued on subsequent days until all criteria were fulfilled satisfactorily. Especially at times of high solar activity, meeting all criteria was challenging, and image series of an as-quiet-as-possible Sun were chosen. The availability of suitable AIA data was limited to a period from February 2011 to May 2018. The limiting factor was poor DEM reconstruction in quiet-Sun regions at the later periods, which is most probably related to the strong degradation of some of the AIA filters. A compilation of the first images of each data set is shown in Fig. \ref{fig:SolarCycleOverview} for the 171, 193, and 335 Å AIA filters. Line-of-sight magnetograms from the Helioseismic and Magnetic Imager \citep[HMI;][]{Schou2012_HMI}, onboard SDO, are included for the beginning of each observation series.

The selected image series were downloaded from the Joint Science Operations Center (JSOC) as full-resolution level 1 data using the available \textit{im\_patch} processing method, which allows images to be pre-cropped to a tracked sub-image surrounding the region of interest before downloading.
We differentially rotated all acquired images of a data set to the time of the first image and then cropped them to the final dimensions of $400\arcsec \times 400\arcsec$ around the center of the solar disk.

Before the DEM reconstruction, clusters of $N \times N$ pixels can be binned in the AIA maps to increase the signal-to-noise. A bin factor $N$ reduces the total number of pixels in an image by $N^2$, increasing the DEM signal-to-noise ratio by about $N$, since the count rate increases by $N^2$, while the Poisson error increases by only $N$. In addition, the number of pixels that deliver no DEM results or unphysical solutions (e.g., because of negative pixel values in some of the high-temperature filters) is significantly reduced. The combination of fewer unsuitable pixels and improved signal-to-noise ratio leads as a whole to fewer missing solutions and more stable DEM results. A further advantage is a reduction in computation time by about $N^2$.

However, a significant drawback is the loss of spatial resolution and consequently a possible reduction in sensitivity to events smaller than the dimensions of the binned pixel. While some binning appears necessary to obtain reliable and robust DEM results, it should be used cautiously, as detecting the smallest possible events is a critical aspect of nanoflare studies. As a compromise, we used the bin-factor $N=4$ for the analysis performed in this study. It strikes a good balance between reduction of the number of unusable DEM pixels on the one hand and the ability to still detect small events on the other hand. A bin-factor of $N=4$ results in a pixel resolution of $2.4\arcsec$ for the AIA images.

As a measure of the solar activity level, we use the 13-month smoothed monthly mean sunspot number derived from the revised sunspot numbers V2.0 \citep{Clette2014}. From now on, we will use the term international sunspot number (ISN) to refer to this 13-month smoothed data unless otherwise mentioned. The data was downloaded from the \citet{sidc}.

\subsection{Differential Emission Measure analysis}
\label{sec:methods_DEM}

We use the regularized inversion algorithm developed by \cite{Hannah2012} to perform the DEM analysis. Required inputs include a map structure of the six selected AIA EUV wavelength channels (94, 131, 171, 193, 211, 335 Å), the edges of the desired temperature bins, and an AIA temperature response function. 
The map structures are created by using each 193 Å image in a data set as a reference and then adding the images of the other five wavelengths with the smallest time separation (typically a few seconds) from the reference image. We use a temperature range of $0.2-8$ MK divided into 20 evenly spaced logarithmic temperature bins because we find that in this interval, the resulting solutions are well constrained by the AIA temperature sensitivity, and the range is wide enough for the quiet-Sun regions under study.
To account for instrument degradation, we always apply the temperature response function for the starting time of the current data set.

Using this approach, we compute the DEM for each time step in each pixel and for all datasets. We then calculate for each pixel the total emission measure (EM) of all temperature bins ($\Delta T$)

\begin{equation}
  EM = \sum DEM(T) \cdot \Delta T
\end{equation}\\
and the DEM-weighted average temperature ($\bar{T}$) \citep[cf.][]{Cheng2012,Vanninathan2015}

\begin{equation}
  \bar{T}=\frac{\sum(DEM(T) \cdot \Delta T \cdot T)}{\sum(DEM(T) \cdot \Delta T)}
\end{equation}\\
for each time step and pixel.

\subsection{Event detection algorithm}
\label{sec:methods_detection}

The detection algorithm searches for events in the derived EM time evolution of each pixel. Three user-specified parameters control crucial steps of the algorithm: the detection interval, the threshold factor, and the combination interval.

First, we compare each time step's EM value against all previous and following values within a set interval. This \emph{detection interval} is defined as the number of 12-second time steps that precede and follow the current time step. If the EM value of the current time step is higher than any other value in the interval, it is marked as a candidate event.

For each marked candidate event, the difference in EM value relative to the minimum EM value since the last candidate event is calculated. The event is accepted if the EM change exceeds a specified threshold. This threshold is defined by multiplying a base threshold, which is calculated for each pixel individually, by a global integer \emph{threshold factor}. To calculate the base threshold, the standard deviation of variations in the EM evolution of each pixel from one time step to the next is used. First, the differences between successive time steps over the entire data set are calculated for each pixel. Then, the standard deviation of these differences is calculated, excluding the upper 10\% of the absolute values. We do this to primarily capture EM variations caused by noise fluctuations and to exclude high values caused by actual events.

Since nanoflares can cover larger areas than a single binned AIA pixel, an algorithm had to be implemented to group detected events in neighboring pixels if they occur within a certain time interval. This interval is defined by the last parameter, the \emph{combination interval}, which specifies the number of 12-second steps in each time direction in which events from adjacent pixels are combined into one event.
Adjacent pixels could either include all surrounding eight pixels or just the ones in the $x$ and $y$ direction (a total of four). Previous nanoflare studies used both approaches, and there is no clear conclusion on which one should be preferred. \cite{Krucker1998} only combined events from the nearest four neighbors with good results, and we also found this approach sufficient for combining events that cover more than one pixel. A higher number of pixels searched for related events mainly increases random encounters and false combinations with little benefit to detecting large-scale events. Therefore, our algorithm only searches for related events in the $x$ and $y$ direction and omits the diagonals. We define the total area of an event as the number of pixels combined in this process.

\subsection{Algorithm parameters}
\label{sec:methods_parameters}

Poorly chosen parameters can favor or limit the detection of certain event energies and thus alter the event distribution, power-law index, and other derived nanoflare properties. However, it has already been shown that finding the perfect event definition is a very difficult task for which there is no clear solution (see \cite{Parnell2004,Parnell2012}). Our experiments with different definitions and parameter settings during the development of the algorithms have shown similar variance in the final results. Thus, we will not attempt to provide a definitive answer as to the best event definition. Instead, our goal was to find reliable parameters that could be used for all datasets and provide comparable results. In addition, our event definition should be suitable for comparison with previous studies.
For this aim, each parameter was tested within a reasonable range, and we then selected the most appropriate one for the final analysis.

For the event detection interval, we tested values of $D=$ 1, 2, 3, and 4. $D=1$ accepts only continuous, uninterrupted increases in the emission measure. During the rising phase, a single low data point splits an event into multiple segments. $D=2$ solves this problem by allowing an event to continue after the interruption. $D=$ 3 or 4 can provide even more visually appealing results but discriminate against smaller events. We have further used a detection interval of $D=2$ to reduce unwanted splitting effects but remain sensitive to the smallest events.
 
We have tested threshold factors of $F=$ 3, 5, 7, and 9 and found that an interval of $F=$ 5 to 7 gives robust results. $F=3$ results in the largest number of events found, many of which are noise, leading to numerous unwanted event combinations. With $F=5$, the number of events detected decreases, and so does the probability of incorrect combinations. With $F=7$, the number of events decreases further but is still usable. At $F=9$, only a small number of events remain due to the high energy threshold. In further analysis, we will use $F=5$ to exclude most of the noise but still remain useful count statistics for the latter frequency analysis.

Lastly, we examined the effects of combination intervals $C=$ 0, 1, 3, 5, 7, and 9. We found satisfactory results for values of $C=$ 3 to 7. $C=3$ appears to be the lower acceptable limit for reliably combining larger events, while $C=5$ results in even less fragmentation of large events into smaller ones and provides more visually appealing results. Increasing this to $C=7$ may be beneficial in combination with a high threshold factor (e.g., $F=7$) but generally produces results similar to $C=5$.
In studies using EIT data, good results have been obtained by either combining only pixels that peak simultaneously or by allowing combinations with the previous or subsequent time step. \cite{Krucker2002} argue about the optimal time interval for combining events in studies with EIT data. They analyzed studies with intervals of $\pm$1 to $\pm$3 min between peaks and concluded that the optimal time interval is closer to $\pm$1 min. Using the frequency of events found in \cite{Krucker1998}, they calculated that the interval of $\pm$3 min has a 45\% chance of producing random encounters in the four adjacent pixels. This chance decreases to 15\% when the interval is only $\pm$1 min.
Since our binned AIA data has a spatial resolution of 2.4 \arcsec compared to the 2.6 \arcsec of the EIT instrument, we can expect similar numbers. With an AIA image cadence of 12 seconds, we must use an interval of $\pm$5 time steps to obtain the same interval of $\pm$1 minute. This factor is what we use in the further analysis.

In summary, the selected parameter set has an event detection interval of $D=2$, a threshold factor of $F = 5$, and an event combination interval of $C = 5$. For more details and illustrations of the different settings, we refer to the thesis of \citet{Purkhart2021}.

\subsection{Event energies}\label{sec:methods_eventenergies}
\label{sec:methods_energy}

We calculate the thermal energy input of the accepted EM enhancements as

\begin{equation}
  E_{th}=3k_BT~A \sqrt{\Delta EM~q~h} \label{equ:Energy}
\end{equation}\\
where $k_B$ is the Boltzmann constant, $T$ the temperature during EM peak derived from the DEM profiles, $\Delta EM$ the increase in emission measure in $\mathrm{cm}^{-5}$, $A$ the total area of the event, $q$ is the filling factor, and $h$ the line-of-sight thickness of the event.

We choose a filling factor of $q=1$ and estimate the line-of-sight thickness by the extent of the event as $h=\sqrt A$, where A is the total event area as defined in sec. \ref{sec:methods_detection}. This definition assumes that all events are loops with a length equal to their width and has been found to give plausible results for studies comparing events over multiple energy ranges \citep{Krucker2002}.

\section{Results}\label{sec:results}

\subsection{Frequency distributions}

\begin{figure*}
\centering
  \includegraphics[width=17cm]{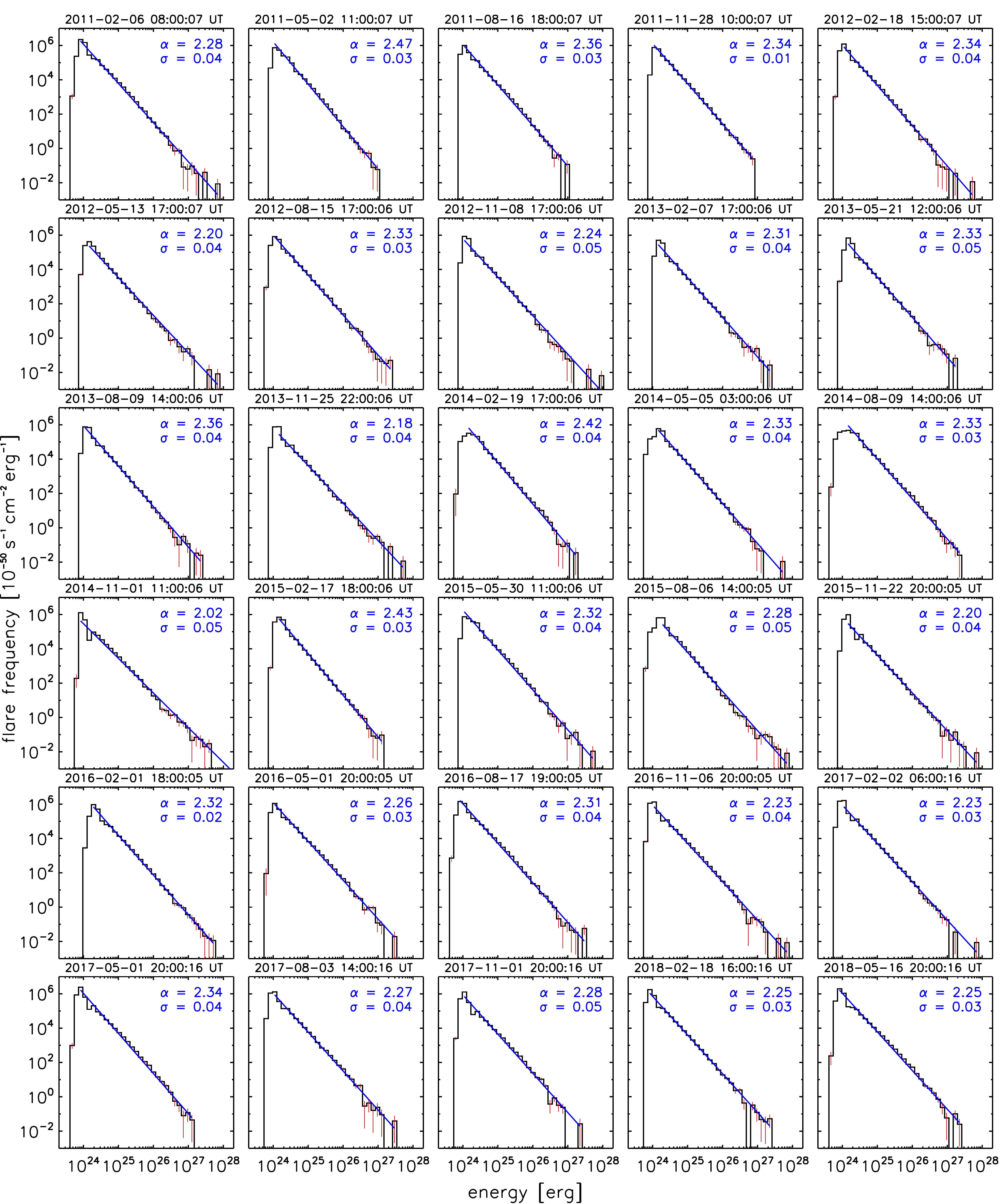}
    \caption{Nanoflare frequency distribution for each data set over the years 2011 to 2018 during solar cycle 24. The start date and time of the corresponding data set are given on top. Events are extracted with the detection interval set to $D=2$, a threshold factor of $F=5$, and a combination interval of $C=5$. A linear fit (blue) without the use of the weights derived from the counting errors (red) was used to extract the power-law index ($\alpha$) and its fitting error ($\sigma$).}
    \label{fig:freqDistNoWeights}
\end{figure*}

\begin{figure*}
\centering
  \includegraphics[width=18cm]{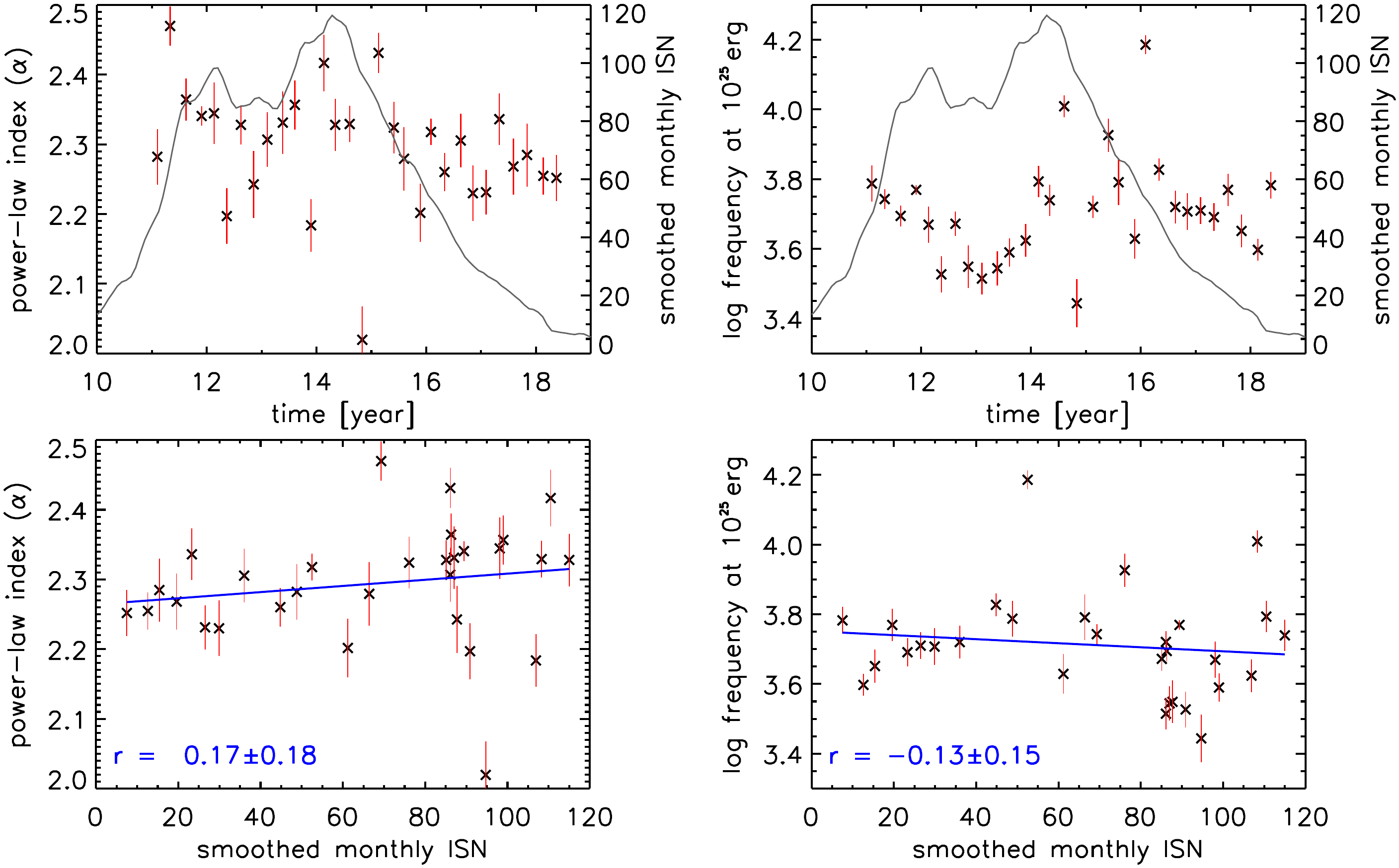}
    \caption{Top: Power-law index $\alpha$ (left) and frequency at $10^{25}$ erg (right) as a function of time along with the 13-month smoothed monthly mean ISN. The fitting error for both parameters is shown as red vertical lines. Bottom: Scatter plots of both fitting parameters against ISN. A linear function (blue) was fitted to the data, yielding the indicated Pearson correlation coefficient $r$ together with the uncertainty range.}
    \label{fig:powerLawLinFitNoWeights_normfactor25LinFitNoWeights}
\end{figure*}

Nanoflare frequency distributions are obtained by constructing a histogram from all event energies in a given data set. We divide the event energies into 50 logarithmic bins uniformly distributed over the energy range from $10^{23}$ to $10^{29}$ erg. The number of events in each group is normalized by the linear energy range covered by that bin, the two-hour observation time, and the area of the observed FOV. Figure \ref{fig:freqDistNoWeights} shows the created nanoflare frequency distributions for all 30 data sets. They are all characterized by a maximum event frequency at about $10^{24}$ erg (cut-off energy) with a sharp turn-over towards the lower energy range. We find that all frequency distributions closely follow a power-law starting at the cut-off energy and continuing to at least $10^{27}$ erg, with some distributions extending to even higher energies up to $10^{29}$ erg.

The measurement errors shown are derived by assuming a Poisson error statistic. Therefore, the error equals the square root of the number of events in that bin. We propagate this Poisson error through the normalization steps and the logarithmization of the histogram to obtain the displayed errors.

The power-law index $\alpha$ and its standard deviation $\sigma$ are derived by a linear fit in log-log space starting at the energy bin with the maximum event frequency and including all non-zero frequencies of the higher energy bins. We omit the indicated errors for the linear fit, resulting in a fit where each energy bin considered is treated with equal weight. Therefore, a small number of large events could significantly affect the steepness of the fit and the resulting power-law index. However, we find that this method results in a linear fit that better reflects the overall slope of the frequency distributions. When Poison errors are used as weights \citep[e.g.][]{Joulin2016}, the fit is almost entirely determined by the lower energy range since these bins contain the majority of counts. Errors from instrument noise, DEM reconstruction, threshold implementation, event combination, and energy calculation are not considered. For most of these errors, the effects on specific energy regions of the histogram are difficult to determine, but they should affect the distribution independently of the count statistics.
 
The individual datasets produce power-law indices in the range of 2.02 to 2.47 with the chosen fitting method. Figure \ref{fig:powerLawLinFitNoWeights_normfactor25LinFitNoWeights} shows these derived power-law indices as a function of time and correlated to the ISN. We fit a linear function and calculate the Pearson correlation coefficient. The uncertainty range of the correlation coefficient is determined utilizing bootstrapping with 10000 resamples. We find that the power-law index shows no significant correlation ($r=0.17\pm0.18$) to the ISN, i.e., no dependence on the solar activity level. 

We also show the same plots for the occurrence frequency at $10^{25}$ erg as determined by the fit. This frequency represents the second fitting parameter ($A$ in Eq. \ref{equ:freqdist}), but with the y-axis shifted to $10^{25}$ erg. This shift has the advantage that it reduces the projection of steepnesses ($\alpha$) variations onto a far-away axis and can, therefore, be used as a better measure of the overall frequency of the observed distributions at a representative event energy. The frequency at this energy level shows a slight decrease during the years 2012 and 2013, but we find no correlation ($r = -0.13\pm 0.15$) to the ISN. Correlations of both parameters to the monthly mean ISN (i.e. without 13-month smoothing) were also calculated but yielded even lower coefficients of $r = 0.10\pm 0.20$ and $r = -0.05\pm 0.15$, respectively.

\begin{figure}
  \resizebox{\hsize}{!}{\includegraphics{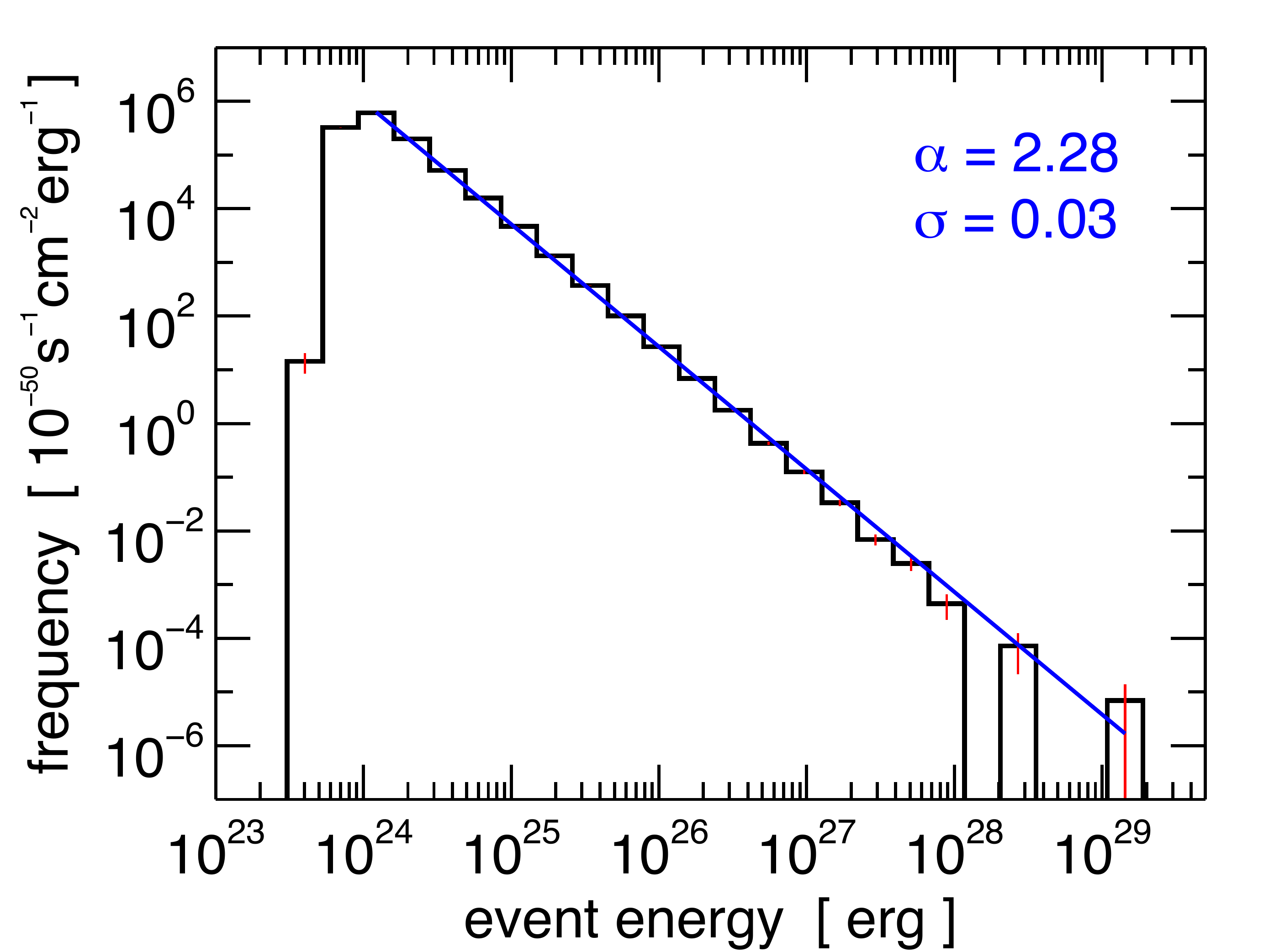}}
  \caption{Combined nanoflare energy distribution (black) derived from all data sets. A linear fit (blue) in log-log space without weights was used to extract the power-law index ($\alpha$) and its fitting error ($\sigma$).}
  \label{fig:freqDistCombined}
\end{figure}

\begin{figure}
  \resizebox{\hsize}{!}{\includegraphics{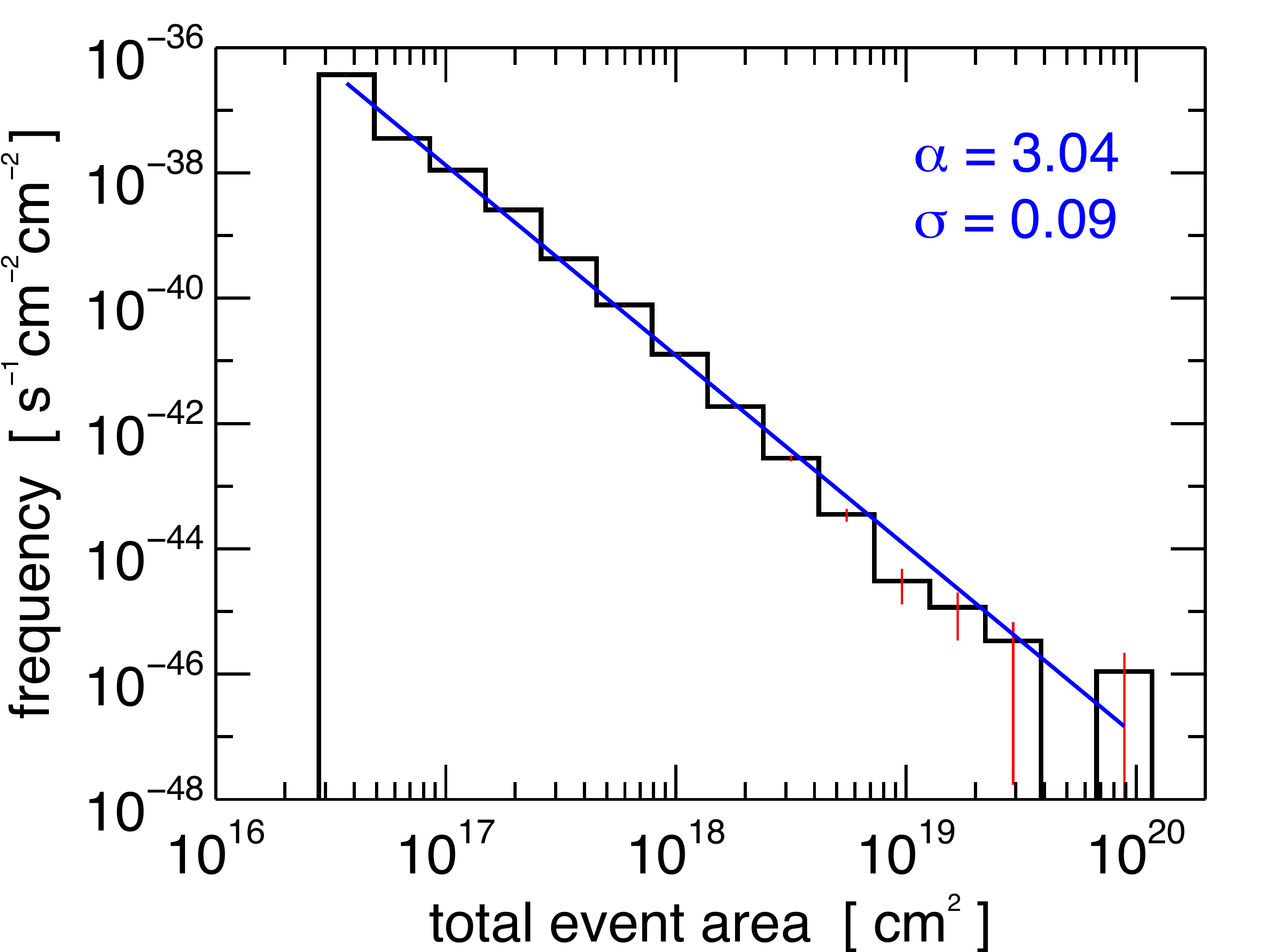}}\caption{Combined nanoflare area distribution (black) derived from all data sets. A linear fit (blue) in log-log space without weights was used to extract the power-law index ($\alpha$) and its fitting error ($\sigma$).}
  \label{fig:freqDistCombinedArea}
\end{figure}

Since there is no dependence on the solar cycle phase, in addition to analyzing the data sets individually, we also combine them to obtain better statistics of the main nanoflare parameters in quiet-Sun regions during solar cycle 24. In Figure \ref{fig:freqDistCombined} we present the event frequency distribution from this combined data set with a total observation time of 60 hours. We use the same parameters as for the individual data sets to extract the events and fit a power-law using a linear fit in log-log space without weights.
The power-law fit shows good agreement with the whole frequency distribution and gives a power-law index of $\alpha = 2.28 \pm 0.03$. The detected events cover more than five orders of magnitude in energy and follow a close power-law distribution from $10^{24}$ to $10^{29}$ erg.
The lower cut-off of the combined frequency distribution at $10^{24}$ erg is expected since all individual distributions also have their cut-off in this range with minimal variation. Individual distributions show a high energy-cutoff of the continuous frequency distribution in the range of $10^{27}$ to $10^{28}$ erg, with only single, separated events found at even higher energies. These high energy events could have significant uncertainties since they may heavily depend on accurate event combinations between many pixels, one of the most challenging steps in the event detection algorithm. It is therefore promising for the developed method that the combined statistics of these large events extend the continuous part of the power-law slope up to $10^{28}$ erg, consistent in steepness with the lower energy part of the frequency distribution that has much better statistics.

Figure \ref{fig:freqDistCombinedArea} shows the distribution of detected total event areas for the combined data sets, with the total event area being the sum of all pixels that were combined into one event. The distribution shows a clear maximum at an event size of $3.1\times 10^{16}\mathrm{~cm^{2}}$ (corresponding to one binned pixel, i.e., $4\times 4$ original AIA pixels) and a continuous decrease in frequency for larger event sizes that follows a power-law over more than three orders of magnitude to almost $10^{20}\mathrm{~cm^2}$ (corresponding to about 3000 binned pixels). For comparison, we note that this maximum area roughly corresponds to the upper range of the H-alpha subflare importance class. We fit a linear function to the distribution and derive a power-law index of $\alpha=3.04 \pm 0.09$. 
Analysis of the individual data sets showed that they all follow a similar power-law distribution with power-law indices between 2.4 and 3.7 and no correlation ($r=0.14\pm0.17$) to the ISN.

\subsection{Energy flux and spatial distributions}

The energy flux is defined as the derived thermal event energy per unit time and area and is therefore given in units of $\mathrm{erg~cm^{-2}~s^{-1}}$. To calculate the mean energy flux of a specific data set, we first sum up the energy from all events observed in that data set and then divide the resulting total energy by the observation duration and the area of the observed FOV. This value can then be compared with the heat flux requirements needed for coronal heating in order to conclude whether the observed nanoflares provide sufficient thermal energy.

\begin{figure}
  \resizebox{\hsize}{!}{\includegraphics{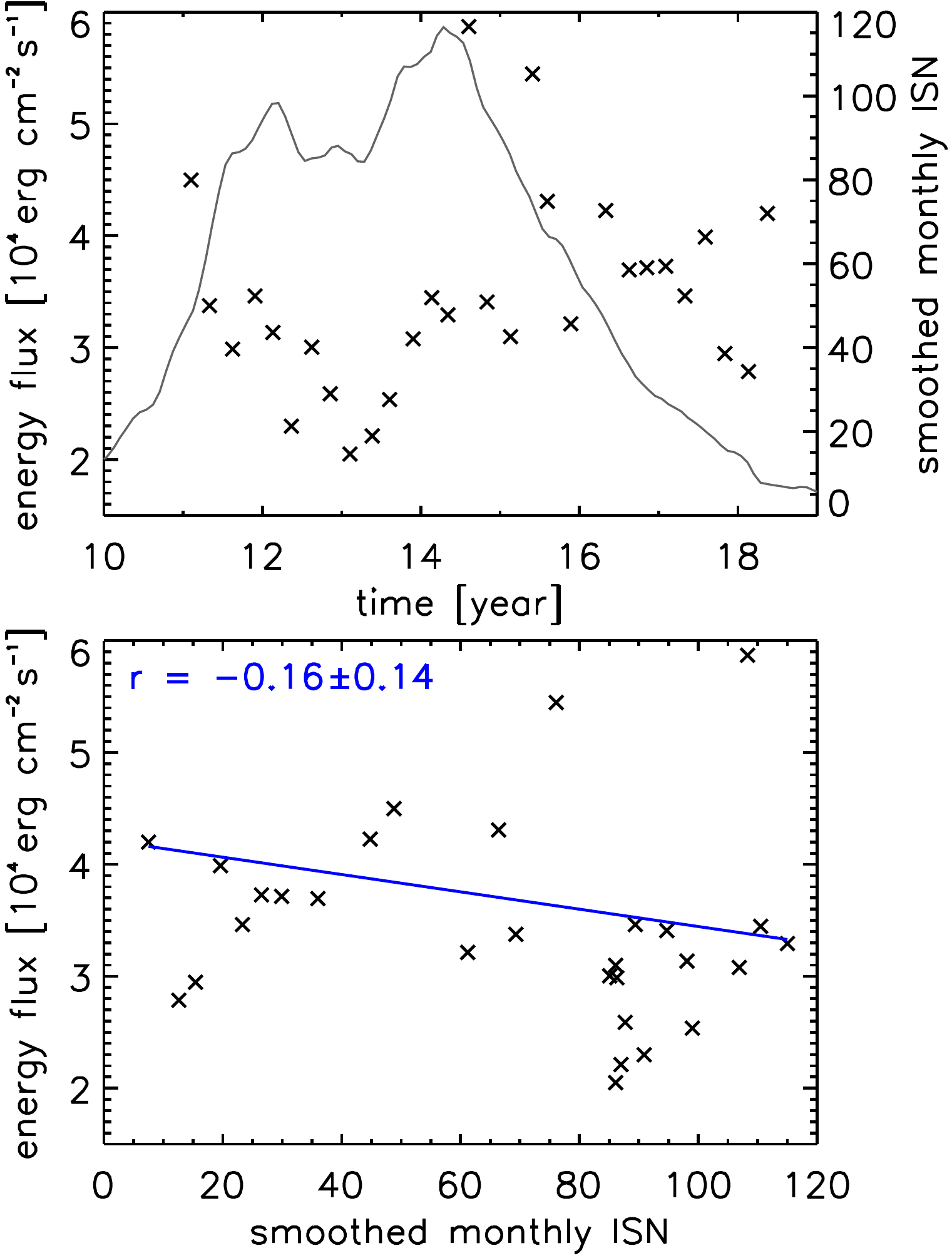}}
  \caption{Top: Mean energy flux from nanoflares as a function of time along with the ISN. Bottom: Scatter plot of mean energy flux against ISN. A linear function (blue) in log-log space was fitted to the data yielding the indicated Pearson correlation coefficient $r$.}
  \label{fig:totalEnergyEvent}
\end{figure}

In Figure \ref{fig:totalEnergyEvent} we show the observed mean energy flux for each data set and their evolution over the solar cycle and the scatter plots against the ISN. The observed mean energy flux remains mostly in the range of $2 \times 10^4 \mathrm{~to~} 6\times 10^4\mathrm{~erg~cm^{-2}~s^{-1}}$. We find a significantly higher mean energy flux (about $1.1\times 10^5\mathrm{~erg~cm^{-2}~s^{-1}}$) in the data set from 2016-02-01. This results in a strong outlier that is not included in the plots. No significant correlation of the mean energy flux to the ISN is found, with a correlation coefficient of $r=-0.16\pm0.14$. However, a dip in mean energy flux during the years 2012 and 2013 coincides with the decreased ISN during this period. This matches the decrease in event frequency at $10^{25}$ erg that was shown in Fig. \ref{fig:powerLawLinFitNoWeights_normfactor25LinFitNoWeights} and is, therefore, likely a result of the slightly decreased overall frequency of the observed events in that timeframe.
From the combined energy distribution in Fig. \ref{fig:freqDistCombined}, we derive a mean energy flux of $(3.7\pm 1.6)\times 10^4\mathrm{~erg~cm^{-2}~s^{-1}}$.

\begin{figure*}
\centering
  \includegraphics[width=16cm]{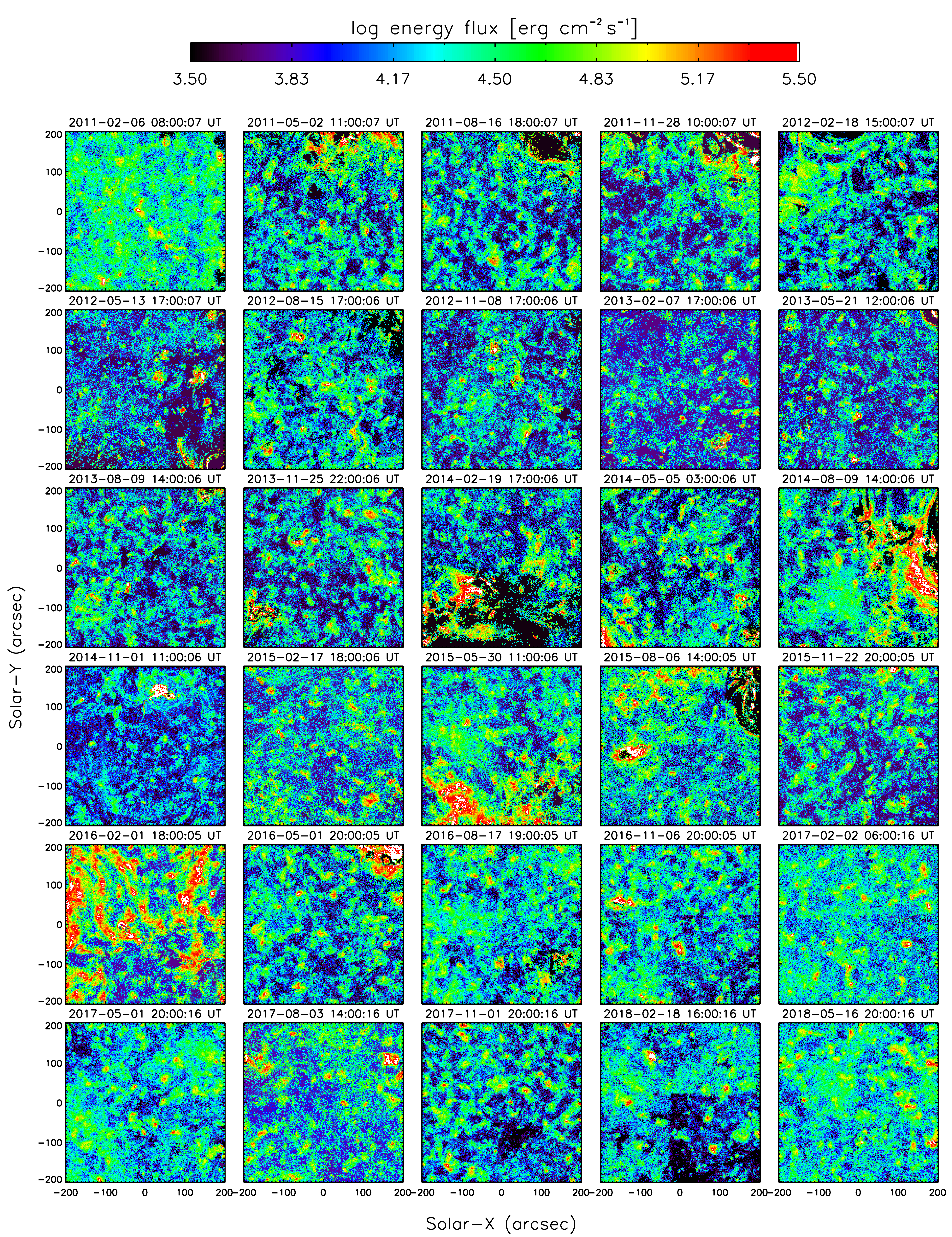}
    \caption{Spatial distribution of the derived energy flux per pixel for all data sets. The beginning of each observation is annotated on top.}
    \label{fig:energyFluxDistribution}
\end{figure*}

To examine the spatial distribution of the thermal energy flux in each dataset, we also calculate the energy flux for each individual pixel by dividing the sum of all event energies detected in that pixel by the observation time and the area covered by just one single pixel. Figure \ref{fig:energyFluxDistribution} shows the obtained spatial distribution of energy flux for all data sets. 
We find that the energy flux is not evenly distributed across the FOV but forms clusters of high energy flux with extended regions of low energy flux in between.

In addition to smaller clusters, some data sets are also characterized by large-scale structures with high energy flux. For example, the 2016-02-01 data set, which was excluded from Fig. \ref{fig:totalEnergyEvent}, shows extensive red bands of high energy flux ($>1.5\times 10^5\mathrm{~erg~cm^{-2}~s^{-1}}$) in the spatial distribution. In other regions, event observations are completely absent and therefore appear black in the images shown. The most prominent example is found in the 2014-02-19 dataset. Interestingly, most areas where no flux was detected are surrounded by areas of high energy flux.
In addition, sharply outlined rectangular regions of reduced energy flux can be seen in the lower right quadrant of some of the later observations, most evident in the 2018-02-18 data set. The reasons for this effect remain unclear.

\begin{figure}
  \resizebox{\hsize}{!}{\includegraphics{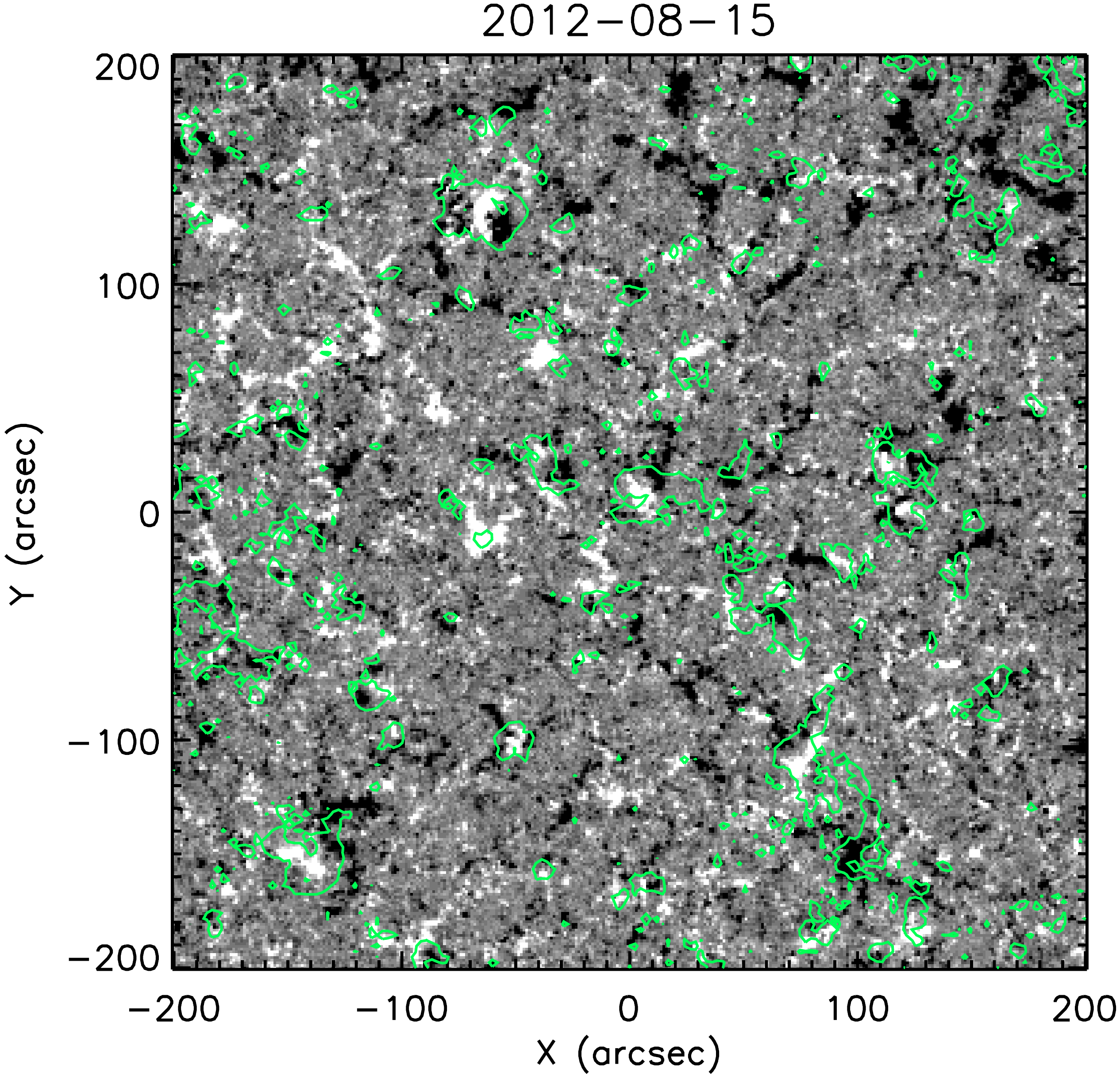}}
  \caption{HMI line-of-sight magnetogram with green contours marking nanoflare regions that exceed an energy flux per pixel of $5\times 10^4\mathrm{~erg~cm^{-2}~s^{-1}}$ over the 2-hr observation time. Magnetogram is saturated at $\pm15$ G.}
  \label{fig:AllCombinedOver6}
\end{figure}

We plot contours of pixels with an energy flux $>5\times 10^4\mathrm{~erg~cm^{-2}~s^{-1}}$ on top of HMI line-of-sight magnetograms in order to study the relationship between high energy flux regions with the underlying photospheric magnetic field. A selected image is shown in Figure \ref{fig:AllCombinedOver6}. We find that the areas with the highest energy flux tend to be located in the magnetic network, with a preference for boundary regions of enhanced, oppositely directed fields.

\begin{figure}
  \resizebox{\hsize}{!}{\includegraphics{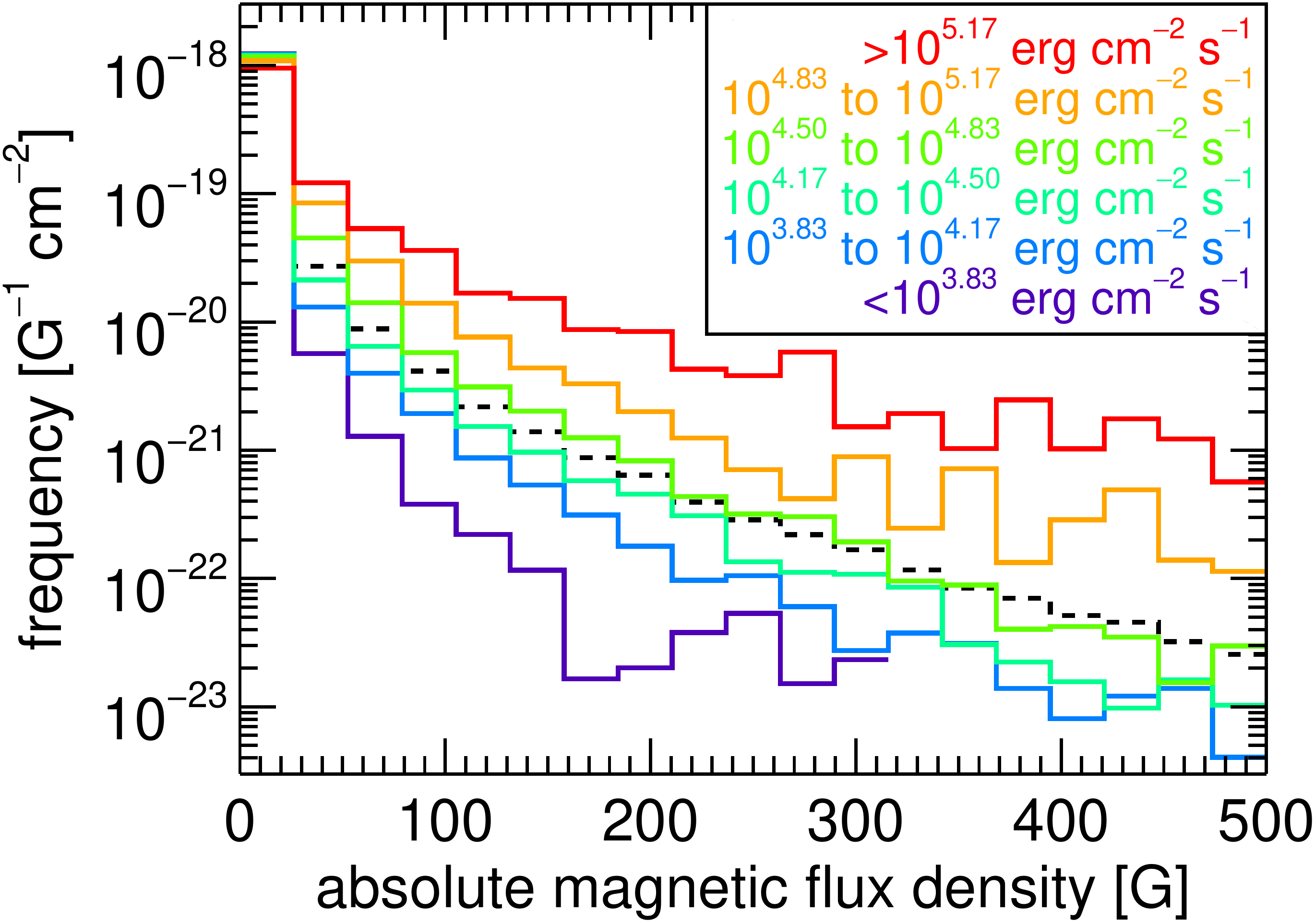}}
  \caption{Frequency distributions of absolute magnetic flux density in areas defined by different energy flux intervals. The whole distribution independent of energy flux is shown as a dashed black line.}
  \label{fig:HMIhisto}
\end{figure}

Figure \ref{fig:HMIhisto} shows frequency distributions of absolute magnetic flux density in pixels corresponding to different energy flux ranges. Information about the spatial distribution of magnetic flux is extracted from the co-registered HMI line-of-sight magnetograms observed at the beginning of each data series. The Figure presents the mean frequency distributions derived from the data of all data sets, excluding pixels with zero energy flux. Frequencies are derived by normalizing each bin by its absolute magnetic flux density range and the area covered by the included pixels. We find that low absolute magnetic flux densities are predominant independent of the chosen energy flux interval. However, regions with higher observed energy flux show significantly increased frequencies of regions with higher absolute magnetic flux density. On the other hand, low energy flux regions show decreased frequencies for high absolute magnetic flux density regions compared to the overall flux density distribution. We find a mean absolute magnetic flux density of 3.1, 4.2, 5.3, 8.0, 14, and 28 G for the logarithmic energy flux intervals of $<3.83$, 3.83 to 4.17, 4.17 to 4.50, 4.50 to 4.83, 4.83 to 5.17, and $>$5.17 $\mathrm{~erg~cm^{-2}~s^{-1}}$.

\begin{figure}
  \resizebox{\hsize}{!}{\includegraphics{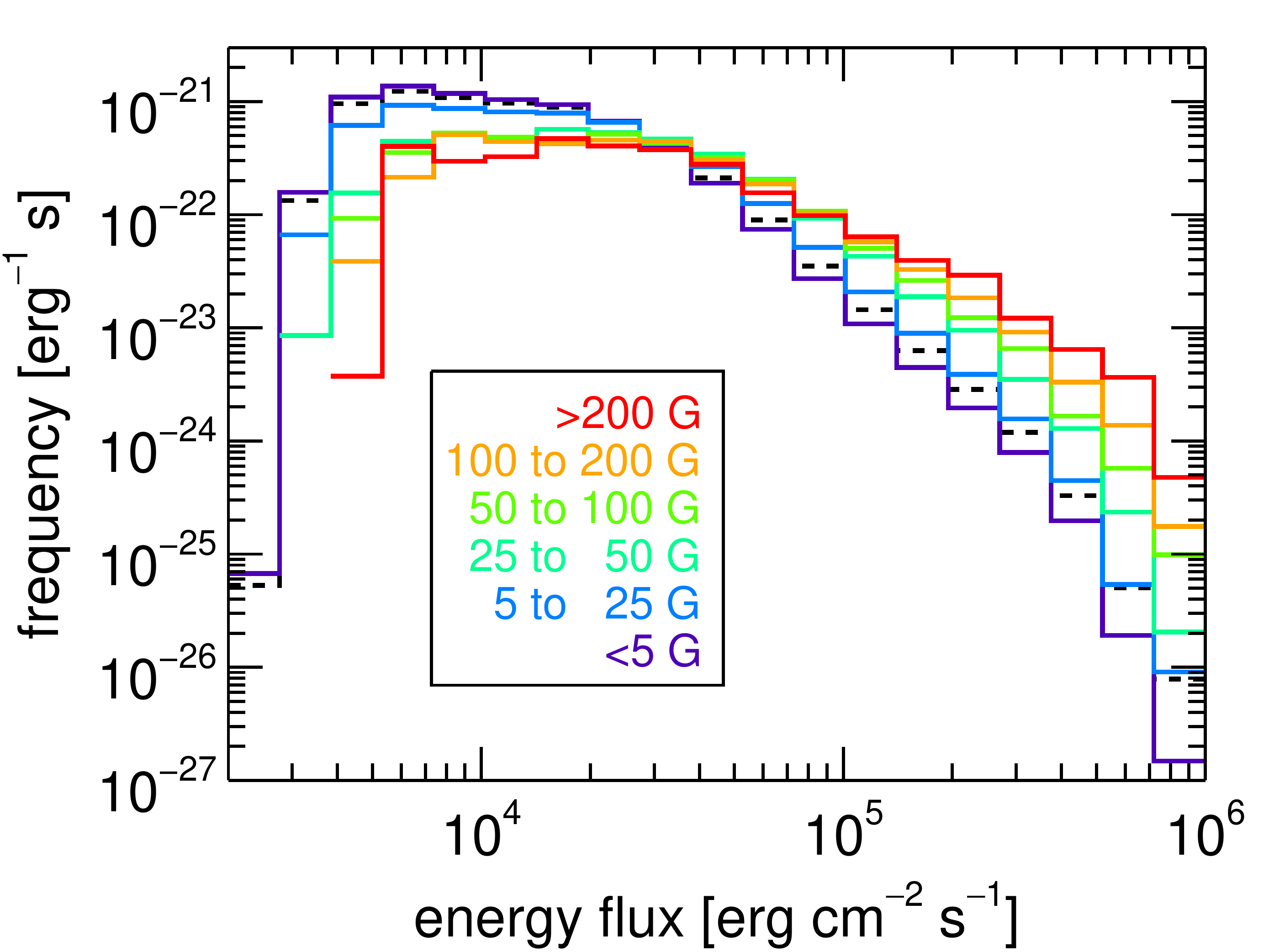}}
  \caption{Frequency distributions of energy flux in areas defined by different absolute magnetic flux density intervals. The whole energy flux distribution independent of underlying magnetic flux density is shown as a dashed black line.}
  \label{fig:FluxHisto}
\end{figure}

Figure \ref{fig:FluxHisto} shows the frequency distribution of energy flux in all pixels from all data sets for different absolute magnetic flux density intervals of the co-registered HMI line-of-sight magnetogram. Frequencies are derived by normalizing each bin by its energy flux range and the area covered by the included pixels. We find that regions with higher absolute magnetic flux density have increased frequencies of high energy flux regions and reduced frequency of low energy flux regions, with a pivot point at about $3\times 10^4\mathrm{~erg~cm^{-2}~s^{-1}}$ for which we find about the same frequency independent of the chosen absolute magnetic flux density interval. We find a mean energy flux of 2.3, 3.3, 5.0, 5.7, 6.7, and 8.9 $\times 10^4\mathrm{~erg~cm^{-2}~s^{-1}}$ in the intervals of $<5$, 5 to 25, 25 to 50, 50 to 100, 100 to 200, and $>200$ G, respectively.

In addition to the energy flux, we also examine the number of events per pixel for each data set. This allows us to study the spatial distribution of nanoflares independent of their energy. We find that a large number of pixels ($>80\%$) have at least one event in all data sets, with some data sets showing activity in nearly $100\%$ of pixels. However, the number of events is not evenly distributed but shows clusters of high activity that are very similar in appearance to the energy flux clusters. We also find square regions of reduced activity in the lower right corner of some data sets that match the previously discussed areas of reduced energy flux. They are characterized by a well-defined edge where the event count suddenly drops for unknown reasons. Therefore, the observed drop in energy flux is possibly due to fewer events in this area.

\section{Discussion}\label{sec:discussion}

\subsection{Power-law}

The individual datasets from the years 2011 to 2018 produce power-law distributions that vary in steepness, with power-law indices in the range of $\alpha = $ 2.02 to 2.47. However, we find no significant correlation between the power-law index and the ISN ($r=0.17\pm 0.18$). This observation is consistent with the avalanche model proposed by \cite{Lu1991} and matches results of the frequency distributions of solar flares in hard X-ray \citep{Crosby1993} and soft X-rays \citep{Veronig2002}.

We also do not find a correlation of the second fitting parameter (represented as the frequency at $10^{25}$ erg in Fig. \ref{fig:powerLawLinFitNoWeights_normfactor25LinFitNoWeights})  with the level of solar activity. These findings imply that in quiet-Sun regions, there is no change in the energy contribution by nanoflares to the coronal heating (and brightness) over the solar cycle. This is in contrast to ARs, where the contribution of regular flares changes in their occurrence frequency. As was shown in the analysis of GOES X-ray flares by \citet{Veronig2002}, the overall frequency varies by more than an order of magnitude over the solar cycle, whereas also in this case, the shape of the distribution ($\alpha$) remains constant.

We have consistently found a power-law distribution of $\alpha>2$ in all data sets, which suggests a dominance of lower energy events in the coronal heating process \citep{Hudson1991}. This is also true for any of the other event detection parameter combinations we applied during the development of the presented method. However, they all rely on other fundamental assumptions that we did not vary. These include, in particular, the thermal energy calculation and filling factor, the scaling of the events line-of-sight depth relative to the detected area ($h=\sqrt{A}$), and the threshold calculation for the event detection.

\cite{Parnell2004} did an extensive study where she investigated the effects of a broader range of assumptions on the observed frequency distribution. She concluded that the exact determination of frequency distributions by direct observations of nanoflares is not possible. From this investigation of 1200 differently tuned event detection and power-law extraction methods, all according to physically sensible assumptions, a wide range of frequency distributions ($\alpha=$ 1.5 to 2.6) was obtained. We note that similar considerations and caveats may apply to any nanoflare frequency studies, including the one presented here. 

\begin{figure}
  \resizebox{\hsize}{!}{\includegraphics{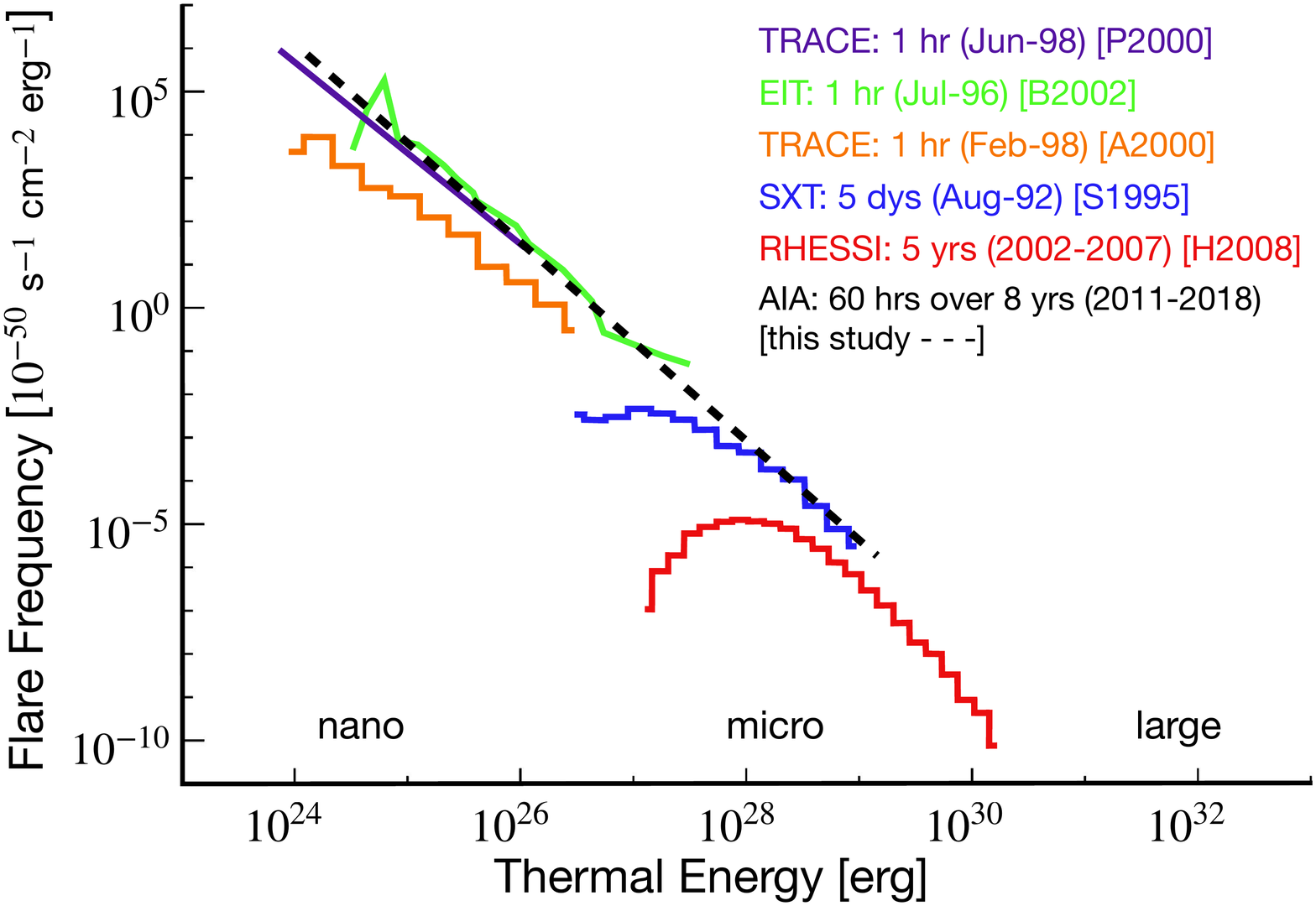}}
  \caption{Comparison of solar flare energy distributions derived by different studies. Shown are EUV nanoflares observed with either SOHO/EIT by \cite{Krucker2002} or TRACE by \cite{Parnell2000} and \cite{Aschwanden2000_1}. Microflares where observed with Yohkoh/SXT by \cite{Shimizu1995} and with RHESSI by \cite{Hannah2008}. The combined AIA nanoflare observations from this study is also included and added by the black dashed line. Figure adapted from \cite{Hannah2011}.}
  \label{fig:HannahCombined}
\end{figure}

Figure \ref{fig:HannahCombined} is adapted from \cite{Hannah2011} and shows our combined nanoflare distribution together with solar nano- and microflare distributions derived by previous studies. 
Our distribution is very similar in steepness to the result from \cite{Krucker2002} who found $\alpha=2.3\pm 0.1$ in their nanoflare studies using EIT data and considering the same height model as was used in this study. The spatial resolution of $2.6\arcsec$ of the EIT instruments is also similar to the $2.4\arcsec$ spatial resolution of the binned AIA pixels used in our study.
A slightly less steep slope was found by \cite{Parnell2000} with a power-law index in the range of $\alpha = $ 2.0 to 2.1 using TRACE data and the same height model.
\cite{Aschwanden2000_1} reported a smaller power-law index of $\alpha = 1.8$ in their nanoflare study of TRACE data.
Our results show an excellent fit to the suggested common power-law distribution of all studies and reach well into the microflare energy range studied by \cite{Shimizu1995} and \cite{Hannah2008}. However, we note that this comparison to microflares is somewhat deceptive as their frequency distributions are derived from soft X-ray observations with different biases and selection effects compared to the EUV nanoflares. Furthermore, microflares occur in active regions, and their occurrence rate is not independent of the solar cycle phase. We have chosen to include the microflare observations to better demonstrate the extensive energy range covered by our study and how the nanoflare energies relate to those of microflares, but we do not directly compare their power-law index and overall frequency. We refer to \citet{Hannah2011} for more detailed discussions on the caveats of combing the different (nano/micro) flare distributions.

What is specific about the present study is the much more extensive range of energies that is covered compared to any previous nanoflare study, spanning over five orders of magnitude. We assume that this is related to the set-up of our study by a) being sensitive also to small and short-lived events (high time cadence, multi-band DEM analysis), b) using different data sets distributed over different phases of the solar cycle, and c) considering in the analysis in total a longer time series than previous nanoflare studies which makes it more likely to catch also the less frequently occurring larger nanoflares. This is also seen in the frequency distributions of the individual data sets (Fig. \ref{fig:freqDistNoWeights}), where all data sets reach to the lower energy cut-off of about $10^{24}$ erg, but not all data sets also show events $>10^{27}$ erg.

\subsection{Energy Flux and contribution to Coronal Heating}

From the individual 30 data sets, we derived mean energy fluxes in the range of $2\times 10^4$ to $6\times 10^4 \mathrm{~erg~cm^{-2}~s^{-1}}$ by averaging the total detected event energy of a data set over the two hour observation time and observed FOV. Our results show no significant correlation between the observed mean energy flux and the ISN. However, a slight tendency of higher energy flux in the second half of the multi-year observation period is visible, as well as a dip in energy flux in the years 2012 and 2013. For the combined distribution of all data sets, we derive a mean energy flux of $(3.7\pm 1.6)\times 10^4\mathrm{~erg~cm^{-2}~s^{-1}}$. This accounts for about $12\%$ of the minimum heating rate of $3\times 10^5 \mathrm{~erg~cm^{-2}~s^{-1}}$ required to heat the solar corona \citep{Withbroe1977}.

\cite{Krucker1998} initially derived a thermal input of $7.1\times 10^4 \mathrm{~erg~cm^{-2}~s^{-1}}$ in their nanoflare study but later corrected these results with the more accurate event depth model ($h=\sqrt{A}$) also used in this study. They arrived at a corrected thermal energy input of about $2.2\times 10^4 \mathrm{~erg~cm^{-2}~s^{-1}}$, which corresponds to about $7\%$ of the minimum heating requirement. \cite{Parnell2004} reported a thermal energy input of $2.9\times 10^4 \mathrm{~erg~cm^{-2}~s^{-1}}$ (about $10\%$) for the matching height model.
Thus, we find similar energy fluxes compared to these previous studies. The generally low values are an inherent problem of this type of nanoflare study and have already been discussed in detail in previous studies \citep[e.g.][]{Krucker2002,Joulin2016}.

Nevertheless, we can assume that the observed frequency distribution is the actual nanoflare frequency distribution and extrapolate how far the distribution would have to continue to even smaller event energies to account for $100\%$ of the heating requirement. If we use the found power-law index of the combined data sets of $\alpha=2.28$, we find that the distribution would have to extend down to about $5\times 10^{20}$ erg, i.e., about 3.5 orders of magnitude lower than the cut-off energy in our study. This requirement moves towards higher energies if our frequency distribution underestimates either the thermal energy release per event (shift to the right) or the overall frequency (shift upwards). Because of our steep slope ($\alpha>2$), an extension of the frequency distribution to larger events (above $10^{29}$ erg) results only in negligible addition to the total energy flux.

However, calculation of the thermal energy input using the EM enhancement at the event peak (see Eq. \ref{equ:Energy}) underestimates the coronal energy input by the observed nanoflares since radiation and conduction losses already reduced the thermal energy during the event. As in nanoflares, the rise time is similar to the decay time, \cite{Krucker2002} estimate that about half of the radiative losses occur before the event peak, which would lead to an underestimate of the thermal energy input by at least a factor of 2. Additional "unobserved" energy is needed for the expansion of chromospheric plasma, where the expansion energy exceeds the thermal energy by about a factor of 3 \citep{Krucker2002}. The reconnection process also produces plasma motions and waves that are not observed. These considerations imply that the nanoflare energy may be up to an order of magnitude larger than the thermal energy derived at the EM peak and demonstrate that the above-given estimate for the needed lower energy extent of the power-law distribution should only be considered a lower limit, with the actually required smallest events being at much higher energies.

\subsection{Spatial Distribution and magnetic characteristics of Events}

We analyzed the spatial distribution of events and energy flux in all data sets and found that they are not distributed homogeneously across the FOV. Instead, they form clusters with increased occurrence and extended areas of reduced activity in between.

We find that most pixels (at least $80\%$) are active in each data set for the used event detection parameter set. This is in contrast to \cite{Parnell2000}, who found that only $16\%$ (for $2\sigma$ events) of the pixels in their quiet-Sun observations contain at least one event. As the main reason, the authors give the short observation time of 15 min and note that \cite{Benz1999} found a significantly higher fraction of active pixels in their 42 min observations. Our high number of active pixels can presumably also be explained by the much longer observation time of 2 hours per data set. Another factor could be the local threshold we set for each pixel instead of one global value. This increases the sensitivity for small events in dimmer pixels where we would otherwise not find any events.

The fraction of active pixels is also dependent on the used threshold factor. Raising the threshold factor to $F=7$ brings the fraction of active pixels in all data sets down below $50\%$ with a large portion of data sets only containing about $20\%$ active pixels. The fraction of active pixels is further decreased to $<20\%$ for all data sets at a threshold factor of $F=9$. We conclude that the number of active pixels will approach a high fraction of the total observed pixels given a long enough observation time and a low enough threshold. However, the combined effect of both factors makes this fraction of active pixels hard to compare between different studies.

Comparisons with HMI line-of-sight magnetograms reveal that the high-activity clusters are located mainly in the magnetic network, preferentially in mixed flux regions of opposite polarities. We find that high energy flux regions show an increased frequency of absolute magnetic flux density in the co-registered HMI magnetograms compared to the overall frequency distribution. The frequency of pixels with absolute magnetic flux density $>200$ G increases by at least an order of magnitude in areas with an energy flux $>10^{5.17}$ compared to the overall distribution.
Furthermore, energy flux frequency distributions of different absolute magnetic flux density intervals show an increase in the frequency of high energy flux pixels in regions with higher absolute magnetic flux density. In regions with absolute magnetic flux density $>200$ G, the occurrence frequency of pixels with energy flux $>2\times 10^5\mathrm{~erg~cm^{-2}~s^{-1}}$ is increased by at least one order of magnitude compared to the overall energy flux distribution. This increase in frequency approaches nearly two orders of magnitude for pixels with energy flux of $10^6\mathrm{~erg~cm^{-2}~s^{-1}}$. At the same time, the frequency of lower magnetic flux density pixels is reduced in those same regions.
The results from both the energy flux and the absolute magnetic flux density distributions show a strong correlation between our detected events and the underlying magnetic field. Together with the finding that they are preferentially located in mixed polarity regions, this makes it likely that the observed nanoflares are indeed magnetic reconnection events.

\section{Conclusion}\label{sec:conclusion} 

In this study, we investigated the frequency distributions of nanoflares and their characteristics in quiet-Sun regions throughout the years 2011 to 2018 to analyze their energy contribution to coronal heating and to study possible changes due to variations in solar activity. We used DEM analysis on the multi-band EUV images from AIA, applying a threshold-based tunable event detection algorithm specifically developed for AIA data characteristics.

For all 30 data sets, we find nanoflare frequency distributions with continuous power-law slopes over multiple orders of magnitude in thermal event energy for all individual data sets.
The frequency distributions calculated from the individual data sets reveal individual power-law indices in the range of $\alpha = $ 2.02 to 2.47. The parameters of the fitted power-law distributions reveal no correlation with the ISN, which indicates that the nanoflare contribution to the quiet-Sun coronal heating does not change over the solar cycle.
The combined frequency distribution of all data sets, with a total observation time of 60 hours at a 12-second temporal resolution, shows a power-law distribution over five orders of magnitude in thermal event energy ($10^{24}$ to $10^{29}$ erg) with a power-law index of $\alpha=2.28\pm 0.03$, indicating that the heat input into the corona is dominated by the lower energy part of the distribution. The mean energy flux derived from the combined data set is $(3.7\pm 1.6)\times 10^4\mathrm{~erg~cm^{-2}~s^{-1}}$, which is about an order of magnitude smaller than the required coronal heat input. The observed frequency distribution would have to continue down to events with a thermal energy of about $5\times 10^{20}$ erg in order to balance the total energy loss of the corona. 

Regions with large energy flux form clusters located preferentially at the boundaries of the magnetic network and show a strong correlation to the underlying absolute magnetic flux density. We found that regions with $>200$ G have an increased frequency of pixels with energy flux $>2\times 10^5\mathrm{~erg~cm^{-2}~s^{-1}}$ by at least one order of magnitude compared to the overall energy flux distribution. These findings suggest that the detected events are small-scale magnetic reconnection events.

\begin{acknowledgements}

AIA and HMI data are courtesy of NASA/SDO and the AIA and HMI science teams. SP and AMV acknowledge the Austrian Science Fund (FWF): I4555-N.
      
\end{acknowledgements}

\bibliographystyle{aa} 
\bibliography{bib-file.bib} 

\end{document}